\documentclass[11pt]{amsart}
\usepackage{times,amssymb,amsmath,float,nicefrac}
\usepackage{euscript}
\usepackage[dvips]{epsfig}

\newtheorem{theorem}{Theorem}

\newcommand{\remove}[1]{}
\newcommand{\new}[1]{#1}

\newtheorem{lemma}{Lemma}

\renewenvironment{proof}[1]{\noindent {\it Proof~:} #1}
{\ \rule{1mm}{2mm}\medskip}

\newcommand\nd{\noindent}

\newcommand\nc\newcommand
\nc\bfa{{\mathbf a}}\nc\bfA{{\mathbf A}}\nc\cA{{\mathcal A}} \nc\bfb{{\mathbf
b}}\nc\bfB{{\mathbf B}}\nc\cB{{\mathcal B}} \nc\bfc{{\mathbf
c}}\nc\bfC{{\mathbf C}}\nc\cC{{\mathcal C}} \nc\bfd{{\mathbf
d}}\nc\bfD{{\mathbf D}}\nc\cD{{\mathcal D}} \nc\bfe{{\mathbf
e}}\nc\bfE{{\mathbf E}}\nc\cE{{\mathcal E}} \nc\bff{{\mathbf
f}}\nc\bfF{{\mathbf F}}\nc\cF{{\mathcal F}} \nc\bfg{{\mathbf
g}}\nc\bfG{{\mathbf G}}\nc\cG{{\mathcal G}} \nc\bfh{{\mathbf
h}}\nc\bfH{{\mathbf H}}\nc\cH{{\mathcal H}} \nc\bfi{{\mathbf
i}}\nc\bfI{{\mathbf I}}\nc\cI{{\mathcal I}} \nc\bfj{{\mathbf
j}}\nc\bfJ{{\mathbf J}}\nc\cJ{{\mathcal J}} \nc\bfk{{\mathbf
k}}\nc\bfK{{\mathbf K}}\nc\cK{{\mathcal K}} \nc\bfl{{\mathbf
l}}\nc\bfL{{\mathbf L}}\nc\cL{{\mathcal L}} \nc\bfm{{\mathbf
m}}\nc\bfM{{\mathbf M}}\nc\cM{{\mathcal M}} \nc\bfn{{\mathbf
n}}\nc\bfN{{\mathbf N}}\nc\cN{{\mathcal N}} \nc\bfo{{\mathbf
o}}\nc\bfO{{\mathbf O}}\nc\cO{{\mathcal O}} \nc\bfp{{\mathbf
p}}\nc\bfP{{\mathbf P}}\nc\cP{{\mathcal P}} \nc\bfq{{\mathbf
q}}\nc\bfQ{{\mathbf Q}}\nc\cQ{{\mathcal Q}} \nc\bfr{{\mathbf
r}}\nc\bfR{{\mathbf R}}\nc\cR{{\mathcal R}} \nc\bfs{{\mathbf
s}}\nc\bfS{{\mathbf S}}\nc\cS{{\mathcal S}} \nc\bft{{\mathbf
t}}\nc\bfT{{\mathbf T}}\nc\cT{{\mathcal T}} \nc\bfu{{\mathbf
u}}\nc\bfU{{\mathbf U}}\nc\cU{{\mathcal U}} \nc\bfv{{\mathbf
v}}\nc\bfV{{\mathbf V}}\nc\cV{{\mathcal V}} \nc\bfw{{\mathbf
w}}\nc\bfW{{\mathbf W}}\nc\cW{{\mathcal W}} \nc\bfx{{\mathbf
x}}\nc\bfX{{\mathbf X}}\nc\cX{{\mathcal X}} \nc\bfy{{\mathbf
y}}\nc\bfY{{\mathbf Y}}\nc\cY{{\mathcal Y}} \nc\bfz{{\mathbf
z}}\nc\bfZ{{\mathbf Z}}\nc\cZ{{\mathcal Z}} \nc\od{{\bar d}}\nc\ow{{\bar
w}}\nc\odelta{{\bar\delta}} \nc\ox{{\bar x}}\nc\oy{{\bar y}}\nc\ou{{\bar u}}
\nc\oh{{\bar h}}

\newcommand\reals{{\mathbb R}}
\newcommand\complexes{{\mathbb C}}

\newcommand\pp{{\mathbb P}}

\nc\dgv{\delta_{\text{\rm GV}}} \nc\rcrit{R_{\text{\rm crit}}}
\nc\Esp{E_{\text{\rm sp}}}
\renewcommand\epsilon{\varepsilon}

\newcommand{\beeq}{\begin{eqnarray*}}
\newcommand{\eneq}{\end{eqnarray*}}

\newcommand{\half}{\nicefrac12}
\font\cyr=wncyr9 scaled 1000\font\cyri=wncyi9 scaled 1000

\begin{document}
\title[LP bounds on codes]{Spectral approach to linear programming
 bounds on codes}\thanks{August 28, 2005, revised November 23, 2005}

\author[A. Barg]{Alexander Barg$^\ast$}\thanks{$^\ast$
Dept. of ECE, University of Maryland,
  College Park, MD 20742.
Supported in part by NSF grants CCR0310961, CCF0515124.}
\author[D. Nogin]{Dmitry Nogin$^\dag$}\thanks{$^\dag$
IPPI RAN, Bol'shoj Karetnyj 19, Moscow 101447, Russia. Supported in part by
NSF grant CCR0310961 and RFFI grant 02-01-22005.}

\begin{abstract} We give new proofs of asymptotic upper bounds
of coding theory obtained within the frame of Delsarte's linear programming
method. The proofs rely on the analysis of eigenvectors of some
finite-dimensional operators related to orthogonal polynomials. The examples
of the method considered in the paper include binary codes, binary
constant-weight codes, spherical codes, and codes in the projective spaces.
\end{abstract}
\maketitle

\subsection*{1. Introduction.} Let $X$ be a compact metric space with
distance function $d$. A code $C$ is a finite subset of $X$. Define the
minimum distance of $C$ as
  $
    d(C)=\min_{\begin{substack}{x,y\in C, x\ne y}\end{substack}}
           d(x,y).
  $
A variety of metric spaces that arise from different applications include the
binary Hamming space, the binary Johnson space, the sphere in $\reals^n$,
real and complex projective spaces, Grassmann manifolds, etc. Estimating the
maximum size of the code with a given value of $d$ is one of the main
problems of coding theory. Let $M$ be the cardinality of $C$. A powerful
technique to bound $M$ above as a function of $d(C)$ that is applicable in a
wide class of metric spaces including all of the aforementioned examples is
Delsarte's linear programming method \cite{del73}. The first such examples to
be considered were the binary Hamming space $H_n=\{0,1\}^n$ and the Johnson
space $J^{n,w}\subset H_n$ which is formed by all the vectors of $H_n$ of
Hamming weight $w,$ with the distance given by the Hamming metric. The best
currently known asymptotic estimates of the size of binary codes and binary
constant weight codes were obtained in McEliece, Rodemich, Rumsey, Welch
\cite{mce77a} and are called the MRRW bounds. Shortly thereafter, Kabatiansky
and Levenshtein \cite{kab78} established an analogous bound for codes on the
unit sphere in $\reals^n$ with Euclidean metric and some related spaces. This
paper also introduced a general approach to bounding the code size in
distance-transitive metric spaces based on harmonic analysis of their
isometry group. This approach was furthered in papers \cite{lev83a,lev98}
which also explored the limits of Delsarte's method.

In this paper we suggest a new proof method for linear programming upper
bounds of coding theory. Our approach, which relies on the analysis of
eigenvectors of some finite-dimensional operators related to orthogonal
polynomials arguably makes some steps of the proofs conceptually more
transparent then those previously known. We also consider some of the main
examples mentioned above, The linear-algebraic ideas that we follow were
introduced in a recent paper by Bachoc \cite{bac05} in which a similar
approach has been taken to establish an asymptotic bound for codes in the
real Grassmann manifold.

\subsection*{2. A bound on the code size}
We assume that $X$ is a distance-transitive space which means that its
isometry group $G$ acts doubly transitively on ordered pairs of points at a
given distance. \new{In this case}\remove{Following \cite{kab78} we will
assume that} the zonal spherical kernels $K_i(\bfx,\bfy)$ associated with
irreducible regular representations of $G$ depend only on the distance
between $\bfx$ and $\bfy.$ In all the examples mentioned above, except for
the Grassmann manifold, $K_i(\bfx,\bfy)$ can be expressed as a univariate
polynomial $p_i(x)$ of degree $i$, where $x=\tau(d)$ is some function of the
distance $d(\bfx,\bfy).$

Let $D$ be the (finite or infinite) set of the possible values of the
distance in $X$. We will assume that $\tau(d(\bfx,\bfy))$ is a monotone
function that sends $D$ to a segment $[a,b].$ For instance, for the Hamming
space, $D=\{0,1,\dots, n\}$ and $\tau$ can be taken the identity function.
For the sphere $S^{n-1}(\reals),$ $D=[0,2].$ In this case it is convenient to
take $\tau(d)=1-d^2/2$ to be the scalar product $(\bfx,\bfy)=\sum_i x_iy_i.$
The invariant measure on $G$ induces a measure $d\mu$ on $[a,b].$ For
instance, for $X=\{0,1\}^n,$ the measure $d\mu$ corresponds to the binomial
probability distribution on $\{0,1,\dots, n\}$, so $\int_Dd\mu=1.$ We will
assume that the last condition holds in general and normalize $\mu$ when this
is not the case.

\new{The kernels $K_i(\bfx,\bfy), i=0,1,\dots,$ are positive
semidefinite which means that $\sum_{\bfx,\bfy\in C}K_i(\bfx,\bfy)\ge 0$ for
any finite set $C\subset X.$ This property together with the fact that
$K_i(\bfx,\bfy)$ can be expressed as a polynomial of one variable gives rise
to the following set of inequalities
   \begin{equation}\label{eq:D}
     \sum_{\bfx,\bfy\in C} p_i(\tau(d(\bfx,\bfy)))\ge 0, \quad
i=0,1,\dots
   \end{equation}
called the Delsarte inequalities in coding theory.}

The function $\tau$ can be chosen in such a way that the polynomials
$p_i,i=0,1,\dots,$ are orthogonal on $[a,b]$ with respect to the
\new{scalar product}
$\langle f,g\rangle=\int fgd\mu.$ Below we denote by $V$ the space
$L_2(d\mu)$ of square-integrable functions on $[a,b].$

We will assume that the polynomials $p_i$ are orthonormal, i.e.,
$\|p_i\|^2=\langle p_i,p_i\rangle=1.$ Note that this implies that $p_0\equiv
1.$ Another assumption used below is that the product $p_ip_j$ for all
$i,j\ge 0$ expands into the basis $\{p_i\}$ with nonnegative coefficients,
i.e.,
   \begin{equation}\label{eq:pijk}
      p_ip_j=\sum_{k} q_{i,j}^k p_k \qquad(q_{i,j}^k\ge 0).
   \end{equation}
This property is \new{again} implied by the fact that the zonal spherical
kernels are positive semidefinite, see \cite{kab78}.

Since the polynomials $\{p_i\}$ are orthogonal, they satisfy a three-term
recurrence \cite{sze75} of the form
  \begin{equation}\label{eq:3term}
    x p_k=\alpha_k p_{k+1}+\beta_k p_k+\gamma_k p_{k-1}
\qquad(k=0,1,\dots;
         p_{-1}=0).
  \end{equation}
Let $P_1=\epsilon p_1,$ where $\epsilon>0$ is some constant. We will write
this recurrence in the form
  \begin{equation}\label{eq:3termP}
    P_1 p_k=a_k p_{k+1}+b_k p_k +c_k p_{k-1},
  \end{equation}
which follows from (\ref{eq:3term}) upon noticing that $P_1$ is a linear
function. By (\ref{eq:pijk}), the coefficients $a_k,b_k,c_k$ are nonnegative.

Let $C\subset X$ be a code of size $M$ and distance $d.$ Denote by
$\Delta(C)=\{\tau(d(\bfx,\bfy)),\, \bfx,\bfy\in C, \bfx\ne \bfy\}$ the set of
values that the function $\tau$ takes on the distances between distinct code
points. Let $\tau_0=\tau(0).$

The main theorem of the linear-programming method asserts the following.
\begin{theorem}\label{thm:lp} \cite{del73,kab78} Let $C\subset X$ be a
code of size $M.$ Let $F(t)=\sum_{i=0}^m F_ip_i(x)$ be a polynomial that
satisfies\\ $(i)$ $F_0>0,F_i\ge 0, i=1,2,\dots,m$;\\ $(ii)$ $F(x)\le 0$ for
$x\in \Delta(C).$

  Then $M\le F(\tau_0 )/F_0.$
\end{theorem}

\new{The proof is obvious because on the one hand, by assumption (ii)
   $$
      \sum_{\bfx,\bfy\in C} F(\tau(d(\bfx,\bfy)))\le MF(\tau_0);
   $$
on the other hand, because of (\ref{eq:D}), assumption (i) and the fact
that $p_0=1,$
   $$
     \sum_{\bfx,\bfy\in C} F(\tau(d(\bfx,\bfy)))=\sum_i
F_i\sum_{\bfx,\bfy}
           p_i(\tau(d(\bfx,\bfy)))\ge F_0M^2.
   $$}

This theorem is equivalent to a duality theorem for a linear programming
problem whose variables are the coefficients of the distance distribution of
the code $C$ and whose constraints are given by the Delsarte inequalities.
For this reason, estimates obtained from this theorem are called the linear
programming bounds. Our objective in this section is to present a new method
of obtaining bounds on $M$ based on this theorem.

We shall use a generic notation $A_k(c_i,b_i,a_i)$ for a tridiagonal matrix
of the form
  $$A_k=\begin{bmatrix}b_0 &a_0 & 0& 0&\dots&0\\
       c_1 &b_1 &a_1& 0&\dots&0\\
        &c_2&b_2&a_2&\dots&0\\
       \dots &\dots &\dots&\dots &\dots&a_{k-1}\\
       0&0 &\dots&\dots &c_k &b_k
     \end{bmatrix}.
  $$
The largest eigenvalue of a square symmetric matrix $M$ will be denoted by
$\lambda_{\max}(M).$

Throughout the paper we use bold letters to denote operators acting on $V$
and regular letters to denote their matrices in the basis $\{p_i\}.$ Let
$V_k$ be the space of polynomials of degree $\le k$ considered as a subspace
of $V$. Let $\bfE_k$ be the orthogonal projection from $V$ to $V_k.$ Consider
the operator
  $$
    \bfS_k=\bfE_k\circ P_1: V_k\to V_k,
  $$
i.e., multiplication by $P_1$ followed by projection on $V_k.$ The argument
that follows relies on the fact that this operator is self-adjoint (with
respect to the bilinear form $\langle\cdot,\cdot\rangle).$ Indeed, both
multiplication by a function and the orthogonal projection are self-adjoint
operators. Therefore, the matrix $S_k=A_k(c_i,b_i,a_i)$ is symmetric. In
other words,
  $$
    a_i=\langle P_1 p_{i},p_{i+1}\rangle=\langle p_{i},P_1
p_{i+1}\rangle
             =c_{i+1}.
  $$

A $p\times p$ matrix $A\ge 0$ (i.e., a matrix with nonnegative entries) is
called irreducible if for any partition of the set of indices
{$\{1,2,\ldots,p\}$} into two disjoint subsets $I$ and $J$, $|I|+|J|=p,$ the
matrix $(a_{i,j})_{i\in I, j\in J}$ is nonzero (in other words, a directed
graph $G$ with vertices $\{1,2,\dots,p\}$ and edges $(i,j)$ whenever
$A_{ij}>0$ is strongly connected). For instance, the matrix $S_k$ is
nonnegative and irreducible.

In the next lemma we collect the properties of irreducible matrices used
below.
\begin{lemma}\label{lemma:pf}
Let $A\ge 0$ be a $p\times p$ irreducible symmetric matrix.

 (a) Its largest
eigenvalue $\lambda_{\max}(A)$ is positive and has multiplicity one. There
exists a vector $y>0$ such that $Ay=\lambda_{\max}(A) y.$

(b) $\lambda_{\max} (A)\le \max_{1\le i\le p} \sum_j A_{ij}.$

(c) For any $y\ne 0$, $\lambda_{\max}(A)\ge \frac{(Ay,y)}{(y,y)}.$

(d) If $0\le B\le A$ for some matrix $B,$ or if $B$ is a principal minor of
$A$, then $|\lambda_{\max}(B)|\le \lambda_{\max}(A).$
\end{lemma}

\medskip\nd
Here claims (a),(b),(d) form a part of the Perron--Frobenius theory (see,
e.g., \cite{Gantm}), and claim (c) is obvious and holds true for any
symmetric matrix.

The suggested method for deriving upper bounds is based on the following
theorem.

\begin{theorem} \label{thm:main}
Let $C\subset X$ be an $(M,d)$ code and let $\rho_k=a_kp_{k+1}(\tau_0
)/p_k(\tau_0 ).$ Then
   $$
     M\le \frac{4\rho_kp_k^2(\tau_0 )}{P_1(\tau_0
)-\lambda_{\max}(S_k)}
   $$
for all $k$ such that $\lambda_{\max}(S_{k-1})\ge P_1(x)$ for all $x\in
\Delta(C).$
\end{theorem}

\begin{proof} Let $g=\sum_{i=1}^k g_i p_i\in V_k.$
Fix some $\rho>0$ (its value to be chosen later). Consider the operator
$\bfT_k:V_k\to V_k$ defined by
  \begin{equation}\label{eq:per}
   \bfT_k g=\bfS_kg-\rho g_kp_k,
  \end{equation}
and let $\theta_k$ be its largest eigenvalue. Recall that $T_k$ is the matrix
of this operator in the basis $\{p_i\}.$ ($T_k$ is the same as $S_k$ except
that $(T_k)_{k+1,k+1}=(S_k)\strut_{k+1,k+1}-\rho.$) We may  ``shift'' the
matrix $T_k$ by a multiple of the identity matrix $I$ to make all of its
elements nonnegative. For instance, we may consider  $T_k+\rho_k I\ge 0.$
Therefore, by Lemma \ref{lemma:pf}(d) we have
$$
\lambda_{\max}(S_{k-1}+\rho I)<\theta_k+\rho<\lambda_{\max}(S_k+\rho I),
$$
whence we get
\begin{equation}\label{*}
\lambda_{\max}(S_{k-1})<\theta_k<\lambda_{\max}(S_k).
\end{equation}
Moreover, the eigenvalue $\theta_k$ is of multiplicity one. Denote by
$f\new{=(f_0,f_1,\dots,f_k)}\in V_k$ the eigenvector that corresponds to it.
By (\ref{eq:per}) we have
  $$
   P_1 f=\theta_k f+\rho f_kp_k+f_ka_k p_{k+1},
  $$
so
  $$
    f=\frac{\rho p_k+a_k p_{k+1}}{P_1-\theta_k} f_k.
  $$
Consider the polynomial $F=(\rho p_k+a_k p_{k+1})f.$ By Lemma
\ref{lemma:pf}(a), $f$ can be chosen to have positive coordinates. Therefore
by (\ref{eq:pijk}), the coefficients of the expansion of $F$ into the basis
$\{p_i\}$ are nonnegative. Next, if $\lambda_{\max}(S_{k-1})\ge P_1(x)$ for
$x\in \Delta(C)$, then by \eqref{*} we have $F(x)\le 0$ for $x\in \Delta(C),$
i.e., $F(x)$ satisfies condition (ii) of Theorem \ref{thm:lp}. Since
multiplication by $f$ is a self-adjoint operator, we compute
 $$
   F_0=\langle(\rho p_k+a_kp_{k+1})f,1\rangle=
      \langle\rho p_k+a_kp_{k+1},f\rangle=\rho f_k>0
 $$
and
  \begin{align*}
    F(\tau_0)&=\frac{(\rho p_k(\tau_0)+a_k p_{k+1}(\tau_0))^2}
{P_1(\tau_0)-\theta_k}f_k
     <\frac{(\rho p_k(\tau_0)+a_k
p_{k+1}(\tau_0))^2}{P_1(\tau_0)-\lambda_{\max}(S_k)}f_k
  \end{align*}
provided that $\lambda_{\max}(S_k)<n.$ Thus,
$$
\frac{F(\tau_0)}{F_0}<\frac{(\rho p_k(\tau_0)+a_k
p_{k+1}(\tau_0))^2}{\rho(P_1(\tau_0)-\lambda_{\max}(S_k))}.
$$
The value of $\rho$ minimizing the left-hand side is $\rho=\rho_k$. The
claimed estimate is obtained by using the polynomial $F=(\rho_k p_k+a_k
p_{k+1})f$ in Theorem \ref{thm:lp}.
\end{proof}

{\em Remark 1.} Note that by Lemma \ref{lemma:pf}(d), the
$\{\lambda_{\max}(S_{k})\}$ form a monotone increasing sequence. Therefore,
the last condition of the theorem holds for all $k$ greater than some value
$k_0.$

Next let us estimate the largest eigenvalue of $S_k$.
\begin{lemma} \label{lemma:lim} Let
$a_{i+1}>a_i, b_{i+1}>b_i, i=0,1,\dots.$ Then for all $s=1,\dots,k+1,$
    $$
      \frac{1}{s}(2(s-1)a_{k-s+1}+sb_{k-s+1})\le \lambda_{\max}(S_k)
                 \le a_{k-1}+\max(a_{k-1}+b_{k-1},b_k).
    $$
\end{lemma}
\begin{proof}
   By Lemma \ref{lemma:pf}(b)
    $$
      \lambda_{\max}(S_k)\le
\max(a_{k-2}+b_{k-1}+a_{k-1},a_{k-1}+b_{k}),
    $$
hence the upper bound. On the other hand, take $y=(0^{k-s+1}1^s)^t$ where $t$
denotes transposition. Then by part (c) of the same lemma,
   $$
     \lambda_{\max}(S_k)\ge
           \frac1s\bigg(2\sum_{p=1}^{s-1}a_{k-p}+\sum_{p=0}^{s-1}
                         b_{k-p}\bigg).
   $$
Since we assumed that the coefficients $a_i,b_i$ are monotone increasing on
$i$ this implies the lower bound.
\end{proof}

{\em Remark 2.} In effect, Lemma \ref{lemma:lim} provides an estimate of the
extremal zero of $p_{k+1}.$ Indeed, consider the operator $\bfX_k=\bfE_k\circ
x: \;V_k\to V_k.$ It is self-adjoint, so its matrix in the basis $\{p_i\}$ is
tridiagonal symmetric and is given by $X_k=A_k(\gamma_i,\beta_i,\alpha_i)$,
where the elements $\alpha_i,\beta_i,\gamma_i$ are the coefficients in the
three-term recurrence (\ref{eq:3term}).

It is well known (e.g., \cite{ism92}) that the spectrum of $X_k$ coincides
with the set of zeros of $p_{k+1}.$ [A proof goes as follows: let
$p_{k+1}(\lambda)=0.$ Consider the action of $\bfX_k$ on the polynomial
$f=p_{k+1}/(\lambda-x)\in V_k:$
   $$
     \lambda f- \bfX_k f=\lambda f- \bfE_k(xf)=
          \bfE_k((\lambda-x)f)=\bfE_k p_{k+1}=0.
   $$
Conversely, if $f\in V_k, f\not\equiv 0,$ and $0=\lambda f-\bfX_k
f=\bfE_k((\lambda-x)f),$ this implies that $(\lambda-x)f$ is a constant
multiple of $p_{k+1}.$ Therefore, $p_{k+1}(x)$ is proportional \footnote{ The
coefficient equals $\alpha_0\alpha_1\dots \alpha_{k-1}$ and can be found
recursively from (\ref{eq:3term}) and the equality $p_0\equiv 1.$} to $\det(x
I_{k+1}-X_k).$] Then the largest zero $x_{k+1}^+$ of $p_{k+1}$ can be found
as $x_{k+1}^+=\lambda_{\max}(X_k)=\max_{\|y\|=1} (X_ky,y),$ or more
concretely as
   $$
     x_{k+1}^+=
        \max_{\|y\|=1}\bigg\{\sum_{i=0}^{k} \beta_i y_i^2+
              2\sum_{i=0}^{k-1} \alpha_i y_iy_{i+1}\bigg\}.
   $$
This formula was first published in \cite[p.580]{lev98} with a different
proof.

\new{We note that the relation between the extremal zero of $p_{k+1}$
and the largest eigenvalue $\lambda_{\max}(X_k)$ makes the task of finding
the zero computationally much easier that the direct approach because of the
existence of very efficient iterative algorithms for the symmetric eigenvalue
problem. This property is helpful for computing linear programming bounds on
codes such as the bounds considered in the next section and other similar
results for codes of moderate or even large length (on the order of several
thousands).}

\subsection*{3. Examples}
In this section we consider a few examples of interest to coding theory.

\subsubsection*{\bf 3.1. \em Binary codes}
Let $X=\{0,1\}^n$ be the binary Hamming space. It is known \cite{del73,kab78}
that the polynomials $p_i$ are given by the (normalized) Krawtchouk
polynomials $\{\tilde K_k(x), k=0,1,\dots,n\}.$ We have $\mu(i)=2^{-n}\binom
ni,$ so the bilinear form can be written as
  $
     \langle f,g\rangle=\sum_{i=0}^n \mu(i) f(i)g(i).
$
Let $C$ be a binary code of length $n,$ size $M$ and minimum Hamming distance
$d=d(C)$. We choose $\tau(k)=k$ to obtain $\Delta(C)\subset
\{d,d+1,\dots,n\}.$ This inclusion may be proper depending on the code $C$,
but we will ignore this and assume that $\Delta(C)=\{d,d+1,\dots,n\}$ since
this assumption can only relax the linear programming bound on $M$.

The polynomials $\tilde K_k$ satisfy a three-term recurrence relation
\cite{sze75}
  \begin{equation}\label{eq:3termK}
    2x \tilde K_k(x)=-\sqrt{(n-k)(k+1)}\tilde K_{k+1}(x)+n \tilde
K_k(x)
           -\sqrt{(n-k+1)k}\tilde K_{k-1}(x),
  \end{equation}
$\tilde K_0=1,$ $\tilde K_i(x)\tilde K_j(x)=\sum_{k} q_{i,j}^k \tilde K_k(x)$
with $q_{i,j}^k\ge 0,$ and
  \begin{equation}\label{eq:zero}
     \tilde K_k(0)=\sqrt{\binom nk}.
  \end{equation}

Choose in (\ref{eq:3termP}) $P_1=\sqrt{n}p_1=n-2x.$ From (\ref{eq:3termK}) we
then obtain $S_k= A_k(a_{i-1},0,a_i),$ where $a_i=\sqrt{(i+1)(n-i)},
i=0,1,\dots,$ or more explicitly,
  $$
    S_k=\left[\begin{array}{cc*4{*6{@{\!}}c}}
          0        &\sqrt n   &0   &\dots&\dots&0\\
          \sqrt n  &0      &\sqrt{2(n-1)} &\dots&\dots&0\\[1mm]
          0&      \sqrt{2(n-1)} &0 &\dots&\dots&0\\
           \dots  &\dots &\dots &\dots&\dots&\dots\\
          \dots&\dots&\dots&0&\sqrt{(k-1)(n-k+2)}&0\\[1mm]
         \dots &\dots&\dots&\sqrt{(k-1)(n-k+2)} &0
&\sqrt{k(n-k+1)}\\[1mm]
         0 &0&\dots&0&\sqrt{k(n-k+1)} &0
  \end{array}\right].
  $$
The monotonicity assumption of Lemma \ref{lemma:lim} clearly holds because
$a_k>a_{k-1}$ as long as $k<n/2.$ Therefore for the largest eigenvalue of
$S_k$ we obtain the following estimate:
            $$\frac{2(s-1)}{s}\sqrt{(k-s+2)(n-k+s-1)}
        \le\lambda_{\max}(S_k)\le 2\sqrt{k(n-k+1)}.$$
Letting $n\to\infty, s\to\infty, s=o(n)$, we obtain the exact asymptotic
behavior of the main term:
    \begin{equation}\label{eq:lh}
\lim_{n\to\infty,\;  k/n\to\tau} \frac{\lambda_{\max}(S_k)}n
                      =2\sqrt{\tau(1-\tau)}.
    \end{equation}
Since $\tau_0=0$ and $\rho_k=n-k,$ the bound of Theorem \ref{thm:main} takes
the form
    \begin{equation}\label{eq:bh}
       M\le \frac{4(n-k)}{n-\lambda_{\max}(S_k)}\binom nk
    \end{equation}
for all $k$ such that $\lambda_{\max}(S_{k-1})\ge P_1(d)=n-2d$. This estimate
together with (\ref{eq:lh}) leads to the following asymptotic result (the
asymptotic MRRW bound for binary codes \cite{mce77a}):
  $$
    \frac 1n \log M\le h(\half-\sqrt{\delta(1-\delta)})(1+o(1)).
  $$
Here $h(x)=x\log_2x-(1-x)\log_2(1-x)$ is the binary entropy function. Indeed,
let $\lim\frac dn=\delta$ and assume that $\delta\le 1/2.$
 We need to choose $k$ so that
$\frac{\lambda_{k-1}}n\ge (1-2\delta)(1+o(1))$ as $n\to \infty$. In the
limit, this amounts to taking $\tau$ that satisfies $2\sqrt{\tau(1-\tau)}\ge
1-2\delta,$ or $\tau\ge \half-\sqrt{\delta(1-\delta)}$. The result now
follows by the Stirling approximation.

\medskip
{\em Remark 3.} Specializing Remark 2 to the case at hand, we observe from
(\ref{eq:3termK}) that
   $$X_k=\half (n I_{k+1}-S_k)=\half A_k(-\sqrt{i(n-i+1)},n,
     -\sqrt{(i+1)(n-i)}).$$
Therefore we obtain the following expression for the largest root of $\tilde
K_{k+1}:$
     \begin{align*}
      x_{k+1}^+=
          \frac n2+\max_{\|y\|=1}
             \sum_{i=0}^{k-1} y_iy_{i+1}\sqrt{(i+1)(n-i)}.
    \end{align*}
This result is originally due to \cite{lev95a}. Although more accurate
estimates of the extremal zeros are available in the literature
\cite{lev98,fos00}, our Lemma \ref{lemma:lim} suffices to compute the correct
value of the main term.

{\em Remark 4.} The bound (\ref{eq:bh}) is close to the previously known
estimates obtained within the frame of Delsarte's method. In particular,
Levenshtein \cite{lev83a,lev98} constructed a sequence of polynomials that
are optimal in the Delsarte problem (with some qualifiers). \new{His
results}\remove{The results of \cite{lev98}} imply that the above estimate
does not improve the known bounds on $M$. The result of \cite{mce77a} is also
of the form similar to (\ref{eq:bh}).

Remarks 2--4, modified appropriately, apply also to the other examples in
this section.

\subsubsection*{\bf 3.2. \em Constant-weight codes}
Now let $X\subset J^{n,w}$ the binary Johnson space, i.e., the set of vectors
in $\{0,1\}^n$ of Hamming weight $w.$ We take $d$ to be the Hamming metric so
that $D=\{0,2,\dots, 2w\}$ and put $\tau(d)=d/2.$ The relevant family of
orthogonal polynomials is given by the Hahn polynomials $H_k(x)$
\cite{del73}. They are orthogonal on $\tau(D)=\{0,1,\dots,w\}$ with respect
to the weight $\mu_{_J}(i)=\frac{\binom wi \binom {n-w} i}{\binom nw}$
according to
   $
     \int H_kH_md\mu_{_J}= \frac{n-2k+1}{n-k+1}\binom nk
\delta_{km}
   $
and satisfy a three-term recurrence
  \begin{multline}\label{eq:3termH}
    (k+1)(w-k)(n-w-k)(n-2k+2)(n-2k+3)H_{k+1}(x)=\\
      (n-2k-1)(n-2k+3)[(n+2)w(n-w)-nk(n-k+1)-(n-2k)(n-2k+2)x]H_k(x)\\
         -(n-2k-1)(n-2k)(w-k+1)(n-w-k+1)(n-k+2) H_{k-1}(x).
  \end{multline}
Note that $\sum_{i=1}^w\mu_{_J}(i)=1$. Let us normalize $H_k$ by setting
$\tilde H_k=\big(\frac{n-2k+1}{n-k+1}\binom nk\big)^{-\half} H_k.$ As above,
we have
  $$
       \tilde H_i(x)\tilde H_j(x)=\sum_{k=0}^w q_{i,j}^k \tilde H_k(x)
          \qquad(q_{i,j}^k\ge 0)
  $$
and
   $$ \tilde H_k(0)=\sqrt{\frac{n-2k+1}{n-k+1}\binom nk}.
   $$
Let us take
      $$
      P_1(x)=(n-1)^{-\half}\tilde H_1(x)=1-\frac{nx}{w(n-w)}.
      $$
Let us write out the matrix of the operator $\bfS_k=\bfE_k\circ P_1$ in the
orthonormal basis. We have $S_k=A_k(a_{i-1},b_i,a_i),$ where the matrix
elements can be computed from (\ref{eq:3termH}). We obtain
  $$a_i=\frac{n(w-i)(n-w-i)}{w(n-w)(n-2i)}
            \sqrt{\frac{(i+1)(n-i+1)}{(n-2i+1)(n-2i-1)}},
   $$
   $$
   b_i=\frac{(n-2w)^2i(n-i+1)}{w(n-w)(n-2i)(n-2i+2)}, \quad i\ge 0.
  $$
Let $C\subset J^{n,w}$ be a code of size $M$ and distance $2d.$ Let us apply
Theorem \ref{thm:main} to bounding $M$ as a function of $d.$ We have
$\tau_0=0, \tilde H_0=1,$
   $$
    \rho_k=a_k\frac{\tilde H_{k+1}(0)}{\tilde H_k(0)}=
         \frac{n(w-k)(n-w-k)(n-k+1)}{w(n-w)(n-2k)(n-2k+1)},
   $$
and $\Delta(C)=\{0,1,\dots,d\}.$ Thus, we obtain the following estimate.
\begin{theorem} \label{thm:bj}
   $$
     M\le \frac{4n(w-k)(n-w-k)}{(1-\lambda_{\max}(S_k))
        w(n-w)(n-2k)}\binom nk
   $$
for all $k$ such that $\lambda_{\max}(S_{k-1})\ge 1-\frac{nd}{w(n-w)}.$
\end{theorem}
Let us find the minimum $k$ that satisfies the required condition. First we
use Lemma \ref{lemma:lim} to compute the asymptotic behavior of
$\lambda_{\max}(S_k).$
\begin{lemma}\label{lemma:lj}
   $$\lim_{\begin{substack}{n\to\infty\\ w/n\to\omega,k/n\to\tau}
        \end{substack}}
           \lambda_{\max}(S_k)=\frac{
             2\omega(1-\omega)+\sqrt{\tau(1-\tau)}}
{\omega(1-\omega)(1+2\sqrt{\tau(1-\tau)})}\sqrt{\tau(1-\tau)}.$$
\end{lemma}
\begin{proof}
Note that for the upper bound in Lemma \ref{lemma:lim} is suffices to prove
that the value $a_i+b_i+a_{i-1}$ grows on $i.$ Letting $\alpha=\frac in,$ we
compute
   $$
a_{i-1}+b_i+a_i=\frac{2(\omega-\alpha)(1-\omega-\alpha)\sqrt{\alpha
   (1-\alpha)}+(1-2\omega)^2\alpha(1-\alpha)}{\omega(1-\omega)
             (1-2\alpha)^2}(1+o(1)).
   $$

$$=\frac{2\omega(1-\omega)\sqrt{\alpha(1-\alpha)}+\alpha(1-\alpha)}
     {\omega(1-\omega)(1+2\sqrt{\alpha(1-\alpha)})}(1+o(1)).
   $$
The main term on the right-hand side of the last expression is a growing
function of $\alpha.$ Indeed, $\sqrt{\alpha(1-\alpha)}$ grows on $\alpha$ for
$\alpha<\half,$ so we only need to check that the function
$t(2\omega(1-\omega)+t)/(1+2t)$ increases on $t$ for $0\le t\le \half$ which
is straightforward. Thus we put $i=k-1$ and obtain for $\lambda_{\max}(S_k)$
an upper bound of the form claimed. Lemma \ref{lemma:lim} also implies a
matching lower bound. Namely, from its proof,
   $$
     \lambda_{\max}(S_k)\ge
    \frac 1s \Big(2\sum_{p=1}^{s-1} a_{k-p}+\sum_{p=0}^{s-1}
b_{k-p}\Big)
  \qquad(s=1,\dots,k+1).
   $$
For large values of the parameters, we can write
   $$
     \lambda_{\max}(S_k)\ge (a_{k-s}+b_{k-s+1}+a_{k-s+1})(1+o(1)).
   $$
The proof is completed by letting $s\to\infty, s=o(n).$
\end{proof}

Let us use this lemma in Theorem \ref{thm:bj}. Assume that $n\to\infty, d=
\delta n.$ The condition on $k$ in this theorem  will be fulfilled for any
$k=\tau/n$ that satisfies
   $$
     \frac{
             2\omega(1-\omega)+\sqrt{\tau(1-\tau)}}
{\omega(1-\omega)(1+2\sqrt{\tau(1-\tau)})}\sqrt{\tau(1-\tau)}
            >1-\frac{\delta}{\omega(1-\omega)}
  $$
or
  $$
\delta>\frac{(\omega-\tau)(1-\omega-\tau)}{1+2\sqrt{\tau(1-\tau)}}.
  $$
We conclude that Theorem \ref{thm:bj} implies the following estimate for an
$(n,M,2\delta n)$ code $C\subset J^{n,w}$ (the asymptotic MRRW bound for
constant weight codes \cite{mce77a}):
  $$
   \frac 1n \log M\le h(\tau)(1+o(1)),
  $$
where $\delta=(\omega-\tau)(1-\omega-\tau)/(1+2\sqrt{\tau(1-\tau)})$.

\subsubsection*{\bf 3.3. \em Spherical codes} Consider codes on the
unit sphere $S^{n-1}$ in $\reals^n$. The polynomials $p_i$ in this case
belong to the family of Gegenbauer polynomials $C_k(x)$
\cite[pp.80ff]{sze75}. We have
   $$
     \int_{-1}^1 C_i(x)  C_j(x)
         (1-x^2)^{\frac{n-3}2}dx=\frac{\binom{n+i-3}{i}}{n+2i-2}
          \omega_n\delta_{i,j},
   $$
where $\omega_n=\frac{\pi\Gamma(n-2)}{2^{n-2}\Gamma^2(\frac{n-2}2)},$ and in
particular for $i=j=0,$ $\int_{-1}^1 (1-x^2)^{\frac{n-3}2}dx=
\omega_n/(n-2).$ We also have $C_k(1)=\binom{n+k-3}k.$

Normalizing the measure, we obtain
$d\mu(x)=\frac{n-2}{\omega_n}(1-x^2)^{(n-3)/2}dx.$ The normalized Gegenbauer
polynomials are then given by
   $$
       \tilde C_k=\sqrt{\frac{n+2k-2}{(n-2) \binom{n+k-3}{k}}}C_k.
   $$
The polynomials $\tilde C_k$ satisfy a three-term recurrence of the form
  $$
     x\tilde C_k(x)=a_k \tilde C_{k+1}(x)+a_{k-1} \tilde C_{k-1}(x),
  $$
where $a_i=\sqrt{\frac{(n+i-2)(i+1)}{(n+2i)(n+2i-2)}}, i=0,\dots,$ and
$\tilde C_{-1}=0, \tilde C_0=1.$ Further, $\tilde C_i\tilde
C_j=\sum_{k}q_{i,j}^k \tilde C_k$ where $q_{i,j}^k\ge 0$ and
   $$
       \tilde C_k(1)=\sqrt{\frac{n+2k-2}{n-2}\binom{n+k-3}{k}}.
   $$
Let $C(n,M,t)$ denote a code in which the angle between any two distinct
vectors $\bfx_i,\bfx_j$ satisfies $\cos
(\bfx_i\hskip-0.9em\widehat{\mbox{\rule{0pt}{1.3ex}\hskip2em}}\hskip-1.1em
,\bfx_j)\linebreak[2]\le t.$ As remarked above, we take $\tau(d)=1-d^2/2.$ We
have $D=[0,2], \tau(D)=[-1,1], \Delta(C)\subset[-1,t],\tau_0=1.$ Choose
$P_1(x)=n^{-\half}\tilde C_1(x)=x,$ then the matrix $S_k$ has the form
$A_k(a_{i-1},0,a_i),$ so
    $$\rho_k=a_k\frac{\tilde C_{k+1}(1)}{\tilde C_k(1)}=
       \frac{n+k-2}{n+2k-2}.
    $$
From Theorem \ref{thm:main} we obtain
\begin{theorem}\label{thm:bs}
        \begin{equation}\label{eq:bs}
     M\le \frac{4}{1-\lambda_{\max}(S_k)}
          \binom{n+k-2}k
   \end{equation}
for all $k$ such that $\lambda_{\max}(S_{k-1})\ge t.$
\end{theorem}
This coincides with the original bound of \cite{kab78}.

\begin{lemma}\label{lemma:lims}
For any $s=2,\dots,k$
    $$
          \frac{2(s-1)}{s}
    \sqrt{\frac{(n+k-s-1)(k-s+2)}{(n+2k-2s+2)(n+2k-2s)}}\le
\lambda_{\max}(S_k)\le 2 \sqrt{\frac{(n+k-3)k}{(n+2k-2)(n+2k-4)}}.
    $$
In particular,
   $$
     \lim_{n\to\infty, \frac kn\to \rho}\frac{\lambda_{\max}(S_k)} n
     =2\frac{\sqrt{\rho(1+\rho)}}{1+2 \rho}.
   $$
\end{lemma}
\begin{proof} We only need to check that $a_i\ge a_{i+1}.$
For $n\ge 5,$
      $$
        a_i^2-a_{i-1}^2=\frac{(n-2)(n-4)}{(n+2i)(n+2i-2)(n+2i-4)}>0,
      $$
so $a_i$ is an increasing function of $i$. The inequalities in the claim now
follow directly from Lemma \ref{lemma:lim}. Letting $s\to\infty, s=o(n)$ and
taking the limit gives the asymptotic behavior of $\lambda_{\max}(S_k).$
\end{proof}

Theorem \ref{thm:bs} and Lemma \ref{lemma:lims} together enable us to recover
the asymptotic bound of \cite{kab78}.
  Namely,
using the Stirling approximation
we obtain
  $$
    \frac1n \log M\le ((1+\rho)\log(1+\rho)-\rho\log\rho)(1+o(1))
  $$
under the condition $t\le \lambda_{\max}(X_{k-1})$ which in the limit of
$n\to\infty, \frac kn\to \rho$ translates
 into $\rho\ge \frac{1-\sqrt{1-t^2}}{2\sqrt{1-t^2}}$.

\subsubsection*{\bf 3.4. \em Codes in projective spaces} A class of
spaces related to the real sphere is given by the projective spaces $\pp
L^{n-1}$ where $L=\reals$ or $\complexes$ of $\mathbb H.$ The zonal spherical
functions in these spaces are given by the Jacobi polynomials
$P_k^{\alpha,\beta}(x)$ \cite{sze75}, where $\alpha=\sigma(n-1)-1,
\beta=\sigma-1,$ and $\sigma=\half, 1,2,$ respectively.

The polynomials $P^{\alpha,\beta}_k(x)$ 
satisfy
  $$
     \int_{-1}^1 P_i^{\alpha,\beta}(x)P_j^{\alpha,\beta}(x)
 (1-x)^\alpha(1+x)^\beta
dx=\frac{2^{\alpha+\beta+1}(k+\alpha)!(k+\beta)!}
          {(2k+\alpha+\beta+1)k!(k+\alpha+\beta)!}\delta_{i,j},
  $$
  $$
    P_k(1)=\binom{k+\alpha}\alpha,
  $$
where by definition $x!=\Gamma(x+1).$ The coefficients of three-term
recurrence (\ref{eq:3term}) have the form
  $$
    \alpha_k=\frac{2(k+1)(k+\alpha+\beta+1)}{(2k+\alpha+\beta+1)
             (2k+\alpha+\beta+2)},\quad
\beta_k=\frac{\beta^2-\alpha^2}{(2k+\alpha+\beta)(2k+\alpha+\beta+2)},
  $$
   $$
     \gamma_k=\frac{2(k+\alpha)(k+\beta)}{(2k+\alpha+\beta)
          (2k+\alpha+\beta+1)}.
   $$
Define the bilinear form on $V$ by $\langle f,g\rangle=\int_{-1}^1 fgd\mu,$
where
   $$
      d\mu(x)= \frac{(\alpha+\beta+1)\binom{\alpha+\beta}{\alpha}}
         {2^{\alpha+\beta+1}}(1-x)^\alpha(1+x)^\beta dx.
   $$
Then the squared norm of $P_k$ is equal to
   $$
     \|P_k^{\alpha,\beta}\|^2=\frac{(\alpha+\beta+1)(\alpha+\beta)!
             (k+\alpha)!(k+\beta)!}
         {(2k+\alpha+\beta+1)\alpha!\beta!k!(k+\alpha+\beta)!}.
   $$
Denote by $\tilde P_k=P_k^{\alpha,\beta}/\|P^{\alpha,\beta}_k\|$ the
normalized Jacobi polynomials.

We will take in (\ref{eq:3termP})
      $$P_1(x)=P_1^{\alpha,\beta}(x)=
\frac12((\alpha+\beta+2)x+\alpha-\beta),$$ then the coefficients of the
recurrence are found to be
   $$
     a_k=\frac{\alpha+\beta+2}{2k+\alpha+\beta+2}
   \sqrt{\frac{(k+\alpha+1)(k+\beta+1)(k+1)(k+\alpha+\beta+1)}
       {(2k+\alpha+\beta+3)(2k+\alpha+\beta+1)}},
  $$
   $$
     b_k=\frac{2(\alpha-\beta)k(k+\alpha+\beta+1)}
     {(2k+\alpha+\beta)(2k+\alpha+\beta+2)},
  $$
and $c_k=a_{k-1}.$

Let $C\subset X$ be a code of size $M$ in which $|(\bfx_i,\bfx_j)|\le t$ for
any two distinct vectors $\bfx_i,\bfx_j.$ We have $D=[0,\sqrt 2],$ so
choosing $\tau(d)=2(1-d^2/2)^2-1$ we obtain
$\tau(D)=[-1,1],\Delta(C)\subset[-1,2t^2-1].$ We compute
$$
     \tilde P_k^2(1)=\frac{(2k+\alpha+\beta+1)}{\alpha+\beta+1}
        \frac{\binom{k+\alpha}\alpha\binom{k+\alpha+\beta}{k}}
             {\binom{k+\beta}{\beta}}.
  $$  $$
   \rho_k=a_k\frac{\tilde P_{k+1}(1)}{\tilde P_k(1)}
         =\frac{(\alpha+\beta+2)(k+\alpha+1)(k+\alpha+\beta+1)}
          {(2k+\alpha+\beta+1)(2k+\alpha+\beta+2)},
  $$
  Using these expressions in Theorem \ref{thm:main} we obtain
\begin{theorem}
    $$
       M\le \frac{4(\alpha+\beta+2)(k+\alpha+1)}
           {(2k+\alpha+\beta+2)(1-\lambda_{\max}(S_k)}
            \frac{\binom{k+\alpha}\alpha\binom{k+\alpha+\beta+1}{k}}
             {\binom{k+\beta}{\beta}}.
    $$
\end{theorem}

Let us use Lemma \ref{lemma:lim} to derive the asymptotic behavior of
$\lambda_{\max}(S_k)$ as $k\to\infty, \alpha=ak,\beta=bk,a>0,b\ge 0.$ We
obtain
  $$
\frac{\lambda_{\max}(S_k)}{k}\to\frac{2\bigl((a+b)\sqrt{(a+1)(b+1)(a+b+1)}
       +(a-b)(a+b+1)\bigr)}{(a+b+2)^2}.
  $$

The condition for Theorem \ref{thm:main} to be applicable is
 \begin{equation}\label{eq:cond}
\lambda_{\max}(S_k)>P_1(2t^2-1)
 =(\alpha+\beta+2)t^2-\beta-1.
  \end{equation}

For instance, let us derive a bound for the case $X={\mathbb P}\reals^{n-1}.$
Letting $k=s n/2,\alpha=(n-3)/2,\beta=-\half,$ we obtain $a=\nicefrac1s,
b=0,$
  $$
   \frac{\lambda_{\max}(S_k)}{k}\to \frac{4(1+s)}{(1+2s)^2}.
  $$
Therefore, for large values of the parameters condition (\ref{eq:cond})
becomes
  $$
   \frac{4(1+s)}{(1+2s)^2}=\frac{t^2}s,
  $$
or $s=\half((1/\sqrt{1-t^2})-1).$ From Theorem \ref{thm:main} we obtain the
asymptotic bound of \cite{kab78} on the code size:
  $$
   \frac 1n \log M\le (1+s)\log(1+s)-s\log s.
  $$
In a similar way we can recover the asymptotic bounds of \cite{kab78} in the
other cases mentioned.

The method presented is a linear-algebraic alternative to the analytic
methods of \cite{mce77a,kab78,lev98}. It is equivalent to them in the sense
that it gives the same asymptotic results, although for finite parameters the
bounds derived by these two approaches generally do not coincide.

\providecommand{\bysame}{\leavevmode\hbox to3em{\hrulefill}\thinspace}

\end{document}